\documentclass[a4paper,11pt]{article}
\usepackage{pos}

\title{From the gravitational waves to the exoplanets: the Research Highlights}
\ShortTitle{Research Highlights}

\author*[a,b,c,d,e]{Maria Giovanna Dainotti}
\author[f,g]{Biagio De Simone}
\author[h]{Nissim Fraija}

\affiliation[a]{Division of Science, National Astronomical Observatory of Japan, 2-21-1 Osawa, Mitaka, Tokyo 181-8588, Japan}
\affiliation[b]{The Graduate University for Advanced Studies (SOKENDAI), Shonankokusaimura, Hayama, Miura District, Kanagawa 240-0115}
\affiliation[c]{Space Science Institute, 4765 Walnut St Ste B, Boulder, CO 80301, USA}
\affiliation[d]{Nevada Center for Astrophysics, University of Nevada, 4505 Maryland Parkway, Las Vegas, NV 89154, USA}
\affiliation[e]{Bay Environmental Institute, P.O. Box 25 Moffett Field, CA, California}
\affiliation[f]{Dipartimento di Fisica, Universit\'a di Salerno, Via Giovanni Paolo II, 132 I-84084 Fisciano (SA), Italy}
\affiliation[g]{INFN, Sezione di Napoli, Gruppo collegato di Salerno, Italy}
\affiliation[h]{Instituto de Astronomía, Universidad Nacional Autónoma de México Circuito Exterior, C.U., A. Postal 70-264, Mexico 04510, Mexico}

\emailAdd{maria.dainotti@nao.ac.jp}
\emailAdd{bdesimone@unisa.it}
\emailAdd{nifraija@astro.unam.mx}

\abstract{\textbf{Abstract}\\
In this Research Highlights, we summarize 31 contributions provided during the Workshop \textit{Multifrequency Behaviour of High Energy Cosmic Sources - XIV}, held in Palermo (Italy) from the 12th to the 17th of June 2023. We will start with the most recent discoveries in the field of gravitational waves (GWs). We will connect this topic to the contributions of Gamma-Ray Bursts (GRBs) associated with GWs and with the Kilonovae (KNe) hunting and, more in general, on GRBs. Continuing on high-energy astrophysics objects, we will delve into Active Galactic Nuclei (AGNs), neutrino astronomy and the study of the primordial universe, both from the space telescopes' observation and from the very recent proposals in terms of cosmological models. From the faraway universe, we will move to the more local scales and discuss the recent observations in Supernova Remnants (SNRs), massive star binaries, globular cluster dynamics, and exoplanets observed by Kepler. 
}

\FullConference{%
  Multifrequency Behaviour of High Energy Cosmic Sources - XIV\\
  12-17 June 2023\\
  Mondello, Palermo, Italy
}

\begin{document}
\maketitle

\section{Introduction}
In this Research Highlights, we include 31 contributions provided during the Workshop \textit{Multifrequency Behaviour of High Energy Cosmic Sources - XIV}, held in Palermo (Italy) from the 12th to the 17th of June 2023, with a brief overview for each topic here included. In Section \ref{sec:GW}, we summarize the observations of the GWs obtained in the LIGO-Virgo-KAGRA (LVK) Collaboration and the implications on the multimessenger astronomy, in particular in the connection of GW with electromagnetic (EM) counterparts. In Section \ref{sec:GRBs}, we discuss the importance of GRB observations both in the context of multimessenger astronomy and in the study of high-energy astrophysical phenomena through the magnetar model. In Section \ref{sec:GRBmissions}, we discuss the current and the future perspectives on the GRB observation missions and we present a proposed space experiment for rapidly detecting KNe. 
In Section \ref{sec:magnetar} we discuss the versatility of the magnetar model in explaining other relevant astrophysical phenomena, in particular, the X-ray bursts. Continuing on the multimessenger astronomy, we summarize in Section \ref{sec:accretion} the importance of the accretion mechanism believed to be the engine for many cosmic events, such as the AGNs. Section \ref{sec:neutrino} summarises the neutrino observations from extragalactic sources. In Section \ref{sec:cosmologyobs}, we discuss the cosmological observations provided by the space telescopes, together with the perspectives provided by the The MIllimetric Sardinia radio Telescope Receiver based on Array of Lumped elements KIDs (MISTRAL), concluding with a quick review of the theoretical aspects on the cosmological structure and Dark Matter (DM) properties. After this discussion, we show the reliability of the Membrane-Universe alternative cosmological model \ref{sec:alternativecosmology}. We then move to the local universe with Section \ref{sec:local}, where we summarize the recent results obtained from the study of SNRs, the binary Wolf–Rayet (WR) star WR140, the globular cluster NGC 6121, and the exoplanets statistics obtained through the Kepler mission.

\section{The era of gravitational waves and perspectives on multimessenger astronomy}\label{sec:GW}
\begin{figure}
    \centering
    \includegraphics[scale=2.5]{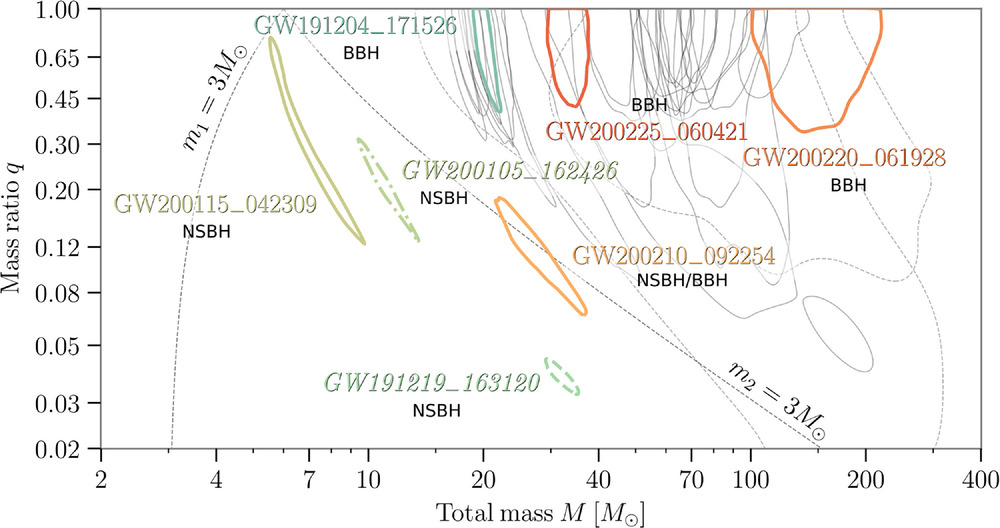}
    \includegraphics[scale=0.2]{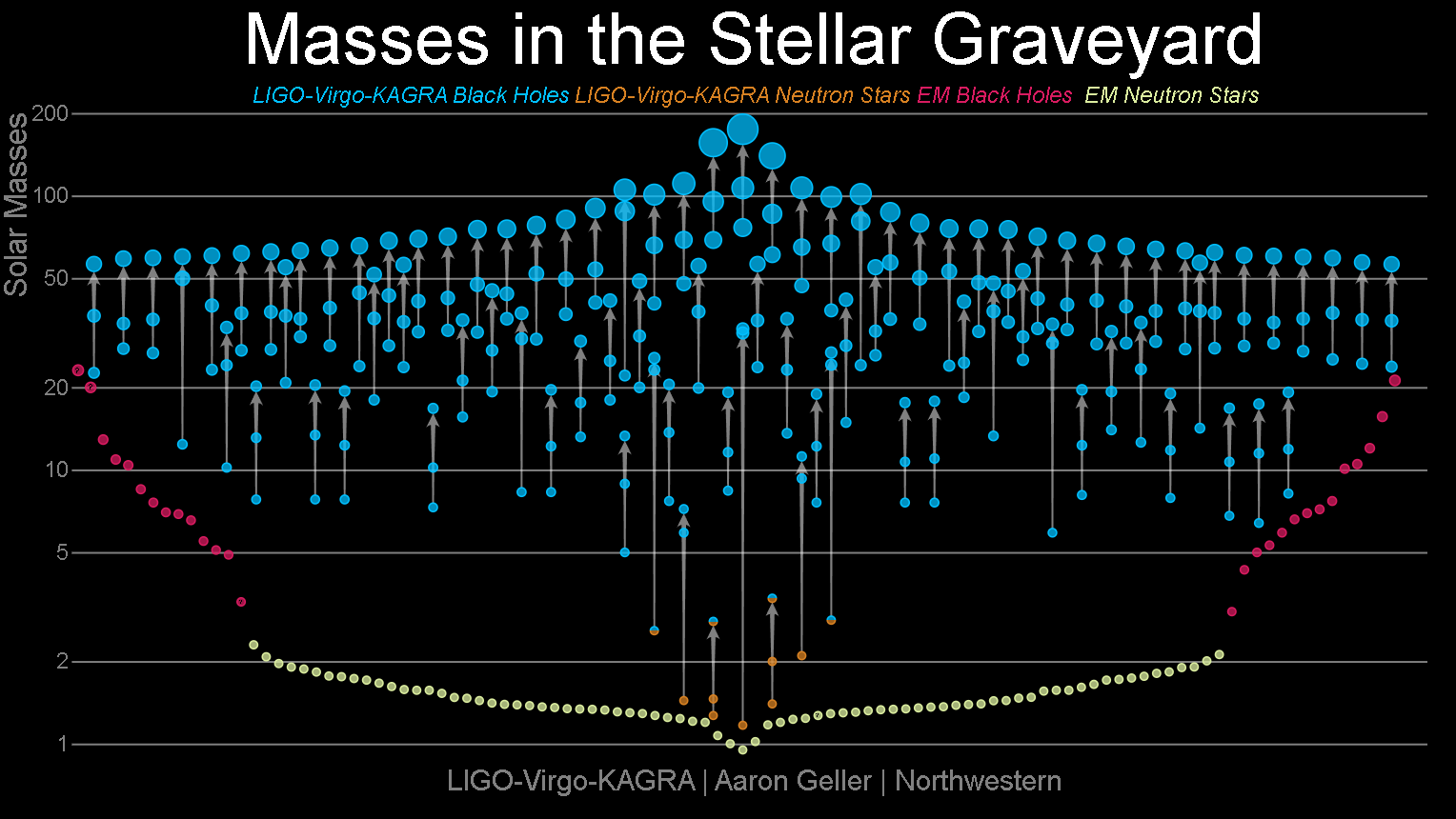}
    \caption{\textbf{Upper panel.} Credible region contours for all O3b candidate with $p>0.5$ and for $GW200105_162426$: the NSBH candidates $GW191219_{163120}$, $GW200105_{162426}$, and $GW200115_{042309}$; the NSBH/low-mass BBH candidate $GW200210_{092254}$, $GW191204_{171526}$, $GW200225_{060421}$, and $GW200220_{061928}$. \textbf{Lower panel.} A diagram that estimates the dead stars' masses as measured by the LVK collaboration \citep{Rosinska2023}.}
    \label{fig:GWs}
\end{figure}

The recent discovery of gravitational waves (GWs) marked the onset of multimessenger astronomy, encompassing a broad spectrum of information derived from astrophysical objects—spanning electromagnetic (EM) signals, neutrino detections, and gravitational wave signatures. The forefront of GW astronomy is built upon the LVK Collaboration. Since its establishment, this network has completed three observation runs: Observation Runs 1, 2, and 3 (O1, O2, and O3, respectively). During O1, the LIGO and Virgo interferometers detected GW150914, an event generated from a binary black hole (BBH) merging. GW150914 represents the first direct observation of GW. O2 witnessed the historic detection of GW170817, the merger of binary neutron stars (BNS), accompanied by associated EM counterparts (the GRB 170817A \citep{GRB170817A} and the KN AT 2017gfo \citep{AT2017gfo}). The O3 cycle, conducted from April 2019 to March 2020, gathered new data but was interrupted by the pandemic. The combination of observations from O1, O2, and O3 provides 90 events collected in the Gravitational-Wave Transient Catalog 1, 2, 2.1, and 3 (GWTC-1, GWTC-2, GWTC-2.1, and GWTC-3, respectively). The candidates in the O3b run (namely, the O3 observations between November 1st, 2019 and March 27th, 2020) are depicted in Figure \ref{fig:GWs}: in the plot, they are presented as 90 $\%$ confidence level contours in the $M-q$ diagram, where $M$ denotes total mass, and $q$ represents the mass ratio of the binary system's components. Events with $p>0.5$ signify high Bayesian odds, determined from signal and noise rates.

The O3 produced several noteworthy candidate events that merit attention \citep{Poggiani2023}:

\begin{itemize}
    \item GW190425: the second BNS merger after the GW170817; 
    \item GW190521: this represents the first observational evidence for intermediate-mass black holes (IMBH), with a mass in the range 100-1000 solar masses ($M_\odot$);
    \item GW190814: this event is peculiar since the secondary object in the merging system could be either a very massive NS or a small mass BH, despite the evidence for it being a NS not reliably strong;
    \item GW200115: this is one of the first NS-BH merging events confirmed.
\end{itemize}

These observations have had a profound impact on various astrophysical and cosmological inquiries. They have been instrumental in understanding the distribution and population of BHs and NSs. They have provided a benchmark to test General Relativity (GR) in the strong field regime, validating its predictions. Moreover, the study of GW events allows for the estimation of the Hubble constant ($H_0$), namely, the expansion rate of the universe observed today. By utilizing data regarding the population of binary black holes (BBHs) and the associated host galaxies to infer the redshift ($z$) of GW events, the value of $H_0$ has been constrained to approximately $68^{+8}_{-6}\,km\,s^{-1}Mpc^{-1}$. Despite meticulous follow-up studies involving EM and neutrino observations, no confirmed counterparts have been identified for the events observed during O3. The LVK Collaboration diligently pursued potential associations with Fast Radio Bursts (FRBs) from the Canadian Hydrogen Intensity Mapping Experiment (CHIME) and GRBs identified by the Fermi and Swift satellites. However, no credible counterpart was identified for the GWs.

The GWTC-3 catalogue has 90 GW detections, BHs and NSs previously constrained through EM observations \citep{Rosinska2023}. The O3 cycle observations confirmed that all the possible pairs of objects in binary mergers could generate a GW emission, and the future observations of cycle O4, which started in May 2023, will shed more light on the field of GW candidates.

Efforts have been dedicated to establishing potential correlations between GW phenomena and optical observations of Core-Collapse Supernovae (CCSNe) as detailed in \citep{Szczepanczyk2023}. The purpose of the work in \citep{Szczepanczyk2023} is to find any connection between GW and CCSNe events through the analysis of the Advanced LIGO and Advanced Virgo detections up to a 30 Megaparsec (Mpc) distance from the Earth. While the study did not ascertain any significant link between a GW signal and a CCSN, it did establish crucial detection limitations. For neutrino-driven explosions, the efficiency of 50 $\%$ in the observations is reached at 8.9 kpc. Additionally, by constraining the mechanism behind CCSNe within the frequency range $50\,Hz-2\,kHz$, \citep{Szczepanczyk2023} outlined the upper limits on the GW energy and luminosity with the values $10^{-4}M_\odot c^{2}$ and $5*10^{-4}M_\odot c^{2}s^{-1}$, respectively. 

\begin{figure}
    \centering
    \includegraphics[scale=1.5]{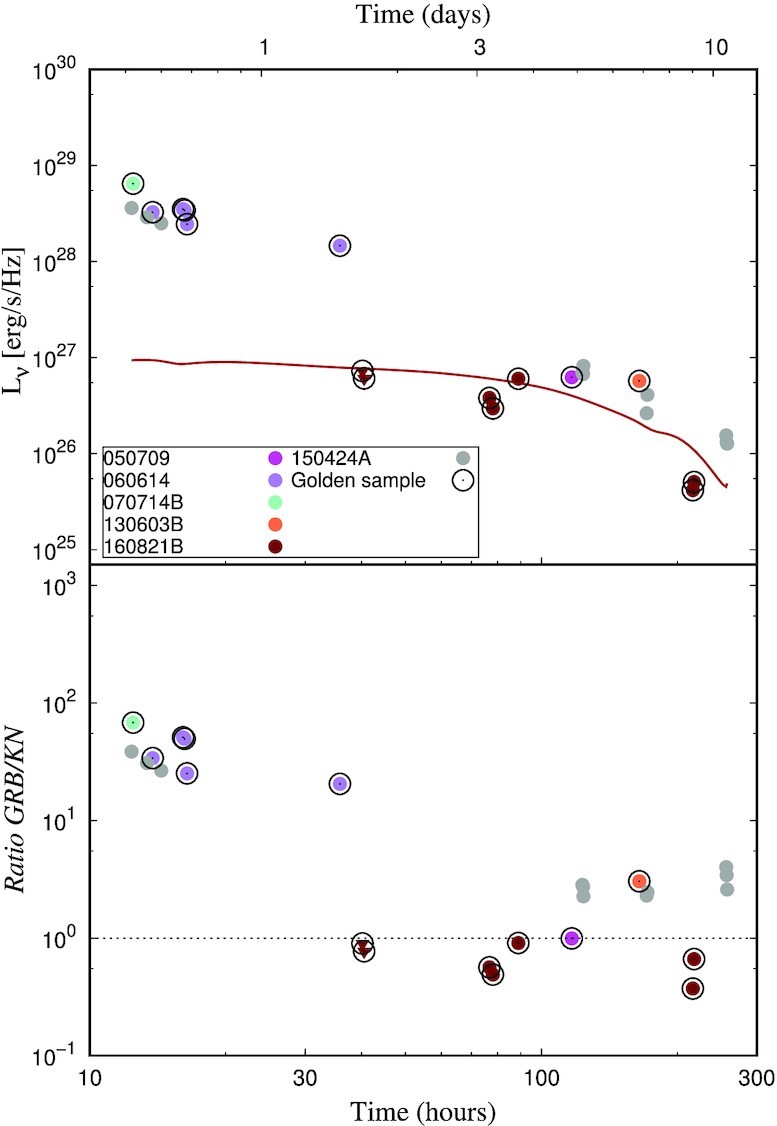}
    \caption{The luminosity of KNe in the red component (wavelength $>900\,nm$) compared with the AT 2017gfo \citep{Rossi2020}.}
    \label{fig:KNe}
\end{figure}

In the domain of multimessenger astronomy, a significant event occurred on August 17, 2017: the detection of the gravitational wave signal GW170817, concurrent with the observations of GRB 170817A \citep{GRB170817A} and the visible counterpart (AT 2017gfo) \citep{AT2017gfo}. This event is solid evidence for detecting a Short GRB (SGRB) progenitor and marks the first confirmed observation of a KN. It also provides crucial insights into the outflows' geometry, the first instance of an off-axis event.
The multimessenger transient GW170817, coupled with KN AT 2017gfo, is a pivotal reference for comprehensively examining KN properties. Before this event, investigations into KNe associated with Short GRBs (SGRBs) had yielded limited promising outcomes. However, the onset of the multimessenger era post-August 17, 2017, facilitated a more expansive exploration of both the blue (UV-optical) and red (optical-near infrared) components of KNe across a broader spectrum of events, thanks to the distinctive properties of AT 2017gfo. Through an analysis of 39 SGRBs' rest-frame light curves (LCs) in the optical/near-infrared, \citep{Rossi2020} concluded that while not all SGRBs are linked with a KN resembling AT 2017gfo, the blue component luminosities span a range of 0.6-17 times the blue luminosity of AT 2017gfo. However, the red component exhibits similarities to AT 2017gfo across all studied KNe, indicating a range of 0.5-3 times the red luminosity of the reference KN. In Figure \ref{fig:KNe}, Rossi et al. (2020) present the GRB luminosity and the GRB/AT2017gfo luminosity ratios concerning time from the mergers in the red component (wavelength $>900,nm$). The GRBs in the golden sample are highlighted with black circles, representing those with well-determined $z$ along with a shallow decay and/or a claimed KN in the existing literature. The red solid line in the top panel denotes the AT2017gfo luminosity at a wavelength of $1600\,nm$.

In the context of multimessenger astronomy utilizing GWs, \citep{PatricelliBernardini} delve into the forthcoming prospects stemming from the O4 run of LVK in conjunction with data from the Fermi, Swift, INTEGRAL, and the Space-based multi-band astronomical Variable Objects Monitor (SVOM) satellites.
Anticipating the outcomes, \citep{PatricelliBernardini} project a range of 1 to 13 Binary Neutron Star (BNS) merger detections annually, a count that could potentially increase by approximately fivefold when considering the potential detection of a BNS using a single interferometer at a relatively low signal-to-noise ratio. They highlight that the collaborative observations involving both GW and EM counterparts promise to offer deeper insights into critical aspects such as the genesis of heavy elements in KNe, the structural formation of jets, the rates and energetic characteristics of SGRBs, and the specific environmental settings where these transients manifest.

\section{The GRBs observations and the connection with their correlations and the magnetar model}\label{sec:GRBs}
Continuing on the analysis of GRBs, but considering the high-z observations, we can state the GRBs are crucial for the following topics:

\begin{itemize}
    \item To investigate the connection between star formation history at high-$z$ ($z>3$) and the GRB rate, since the production of Long GRBs (LGRBs) is associated with a given threshold in the metallicity values \citep{Vergani} and they seem to prefer environments characterized by a metallicity of $12+\log_{10}(O/H)<8.6$ \citep{GhirlandaSalvaterra};
    \item To track the dust evolution through cosmic times and the circumburst environments for GRBs;
    \item To characterize the reionization epoch, given that the GRBs are observed up to $z=9.4$ \citep{Cucchiara2011};
    \item To detect and find population III stars, which can be the progenitors of GRBs;
    \item To investigate the properties of the faint GRB host galaxies that would otherwise remain hidden.
\end{itemize}

\begin{figure}
    \centering
    \includegraphics[scale=0.63]{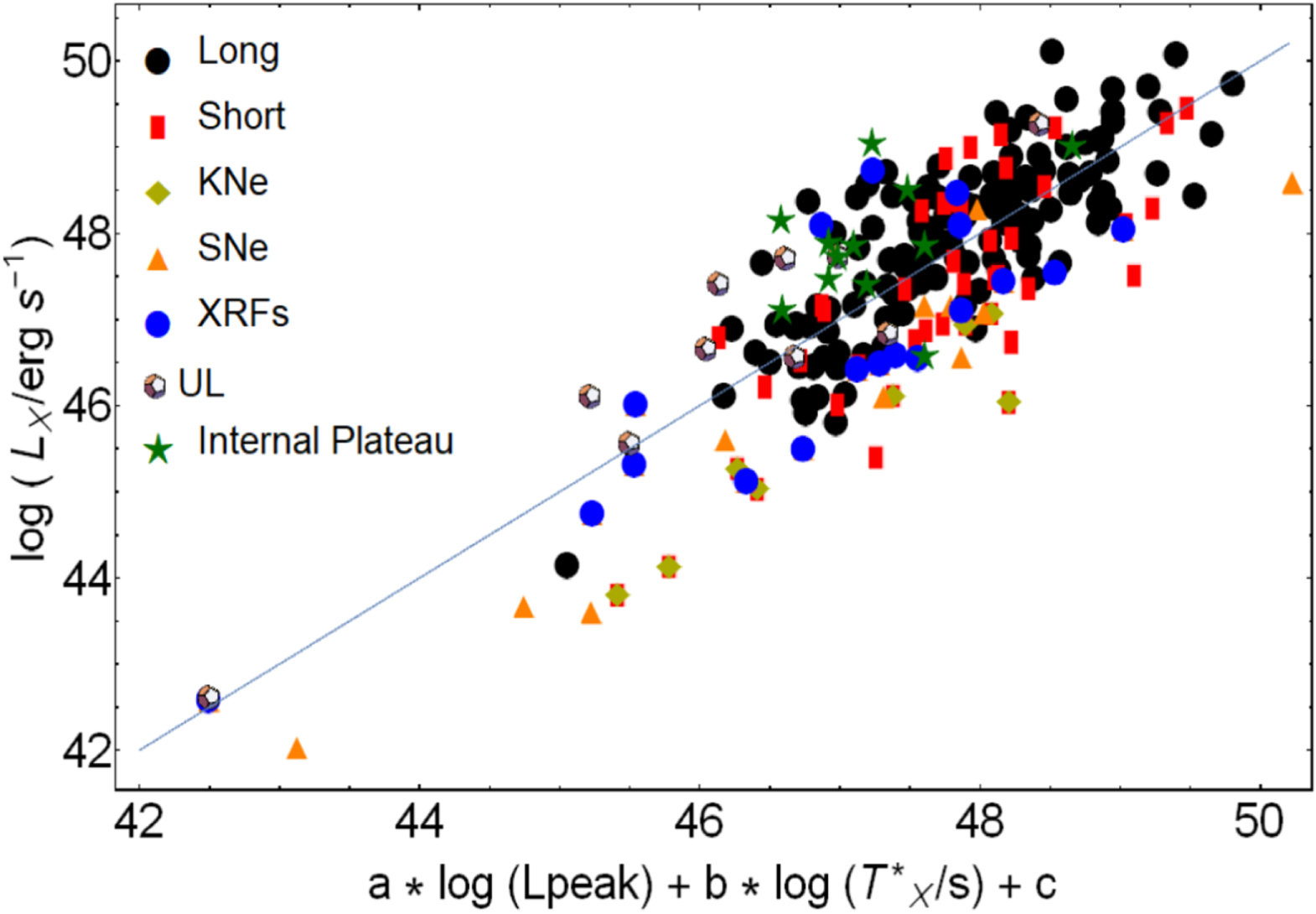}
    \includegraphics[scale=1]{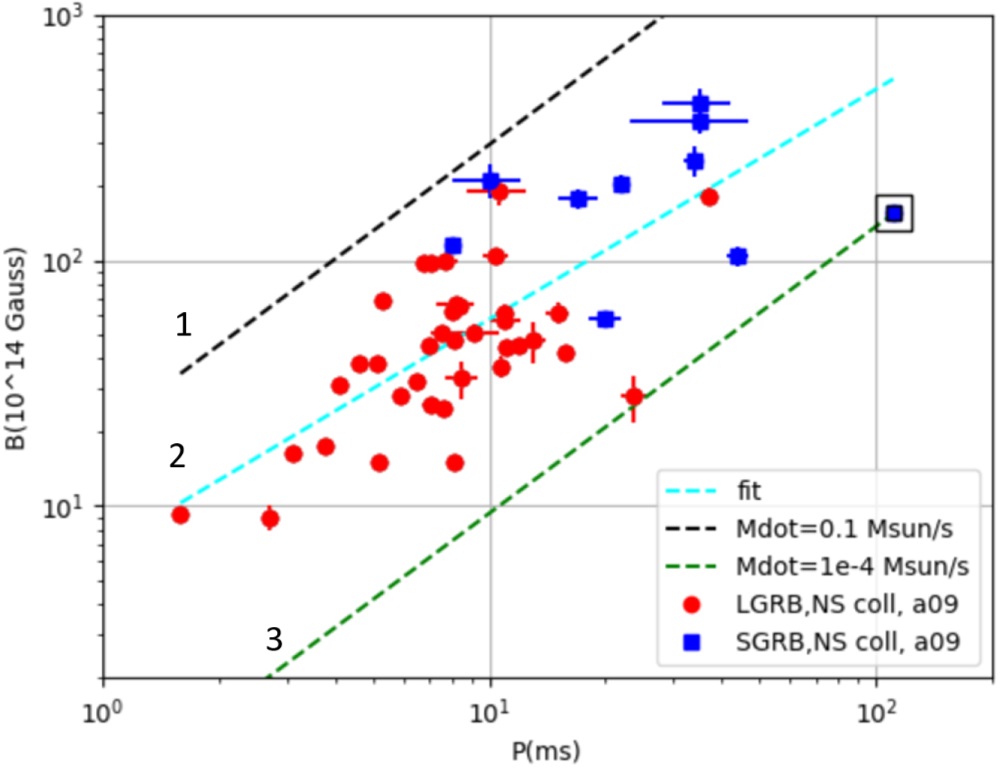}
    \caption{\textbf{Upper panel.} The projection in 2 dimensions of the fundamental plane relation for 222 GRBs. The plot includes the following classes: LGRBs (black dots), SGRBs (red rectangles), KN-SGRBs (green rhombuses), SN-LGRBs (orange triangles), XRFs (blue dots), ULGRBs (dodecahedrons), and GRBs with internal plateaus (green stars). This plot is taken from \citep{Dainotti2020platinum}. \textbf{Lower panel.} The $P-B$ diagram reported in \citep{Stratta2018}. The lines are referred to the expected $P-B$ relations from accreting NSs with an accretion rate of $0.1$ (line 1) and $10^{-4}M_{\odot}/sec$ (line 3); line 2 is the best fit for the $P-B$ relation.}
    \label{fig:Stratta2018}
\end{figure}

We now discuss some relevant features of the GRB LCs. Notably, within 42 $\%$ of GRBs observed via the Swift-XRT telescope, a distinct feature known as the \textit{plateau emission} manifests. This plateau is the flattening of the GRB LC between the prompt emission (observed in the $\gamma$, X-ray wavelengths, and sometimes in the optical) and the late afterglow of the LC (observed in X-rays, optical, and more rarely in the radio wavelengths). This particular feature is present in several relevant astrophysical correlations, namely, the end of plateau luminosity-time relation ($L_X-T^{*}_X$) \citep{Dainotti2008,Dainotti2013,Dainotti2020,Dainotti2022ApJS,Dainotti2022MNRAS,Dainotti2022PASJ} and the fundamental plane relation \citep{Dainotti2016,Dainotti2017,Dainotti2020,Dainotti2022ApJS}, that involves the end of plateau luminosity and time together with the prompt peak luminosity ($L_{peak}-L_X-T^{*}_X$). The fundamental plane relation holds for different classes of GRBs, such as LGRBs, SGRBs, SGRBs associated with KNe (KN-SGRBs), LGRBs associated with Supernovae (SN-LGRBs), X-ray flashes (XRFs), Ultra-Long GRBs (ULGRBs), and GRBs with internal plateau. A plot of the fundamental plane relation can be seen in the upper panel of Figure \ref{fig:Stratta2018}. This relation can be used to promote GRBs to standardizable candles and leverage them as cosmological tools \citep{Dainotti2022MNRAS,Dainotti2022PASJ}.  
It is crucial to address the standard candle from a physical point of view since it can be used as a cosmological tool \citep{Bargiacchi2023a,Bargiacchi2023b,Cao2022a,Cao2022b,Dainotti2022MNRAS,DainottiBargiacchi2023,Lenart2023}.
To this end, the plateau emission observed in GRBs finds a natural explanatory framework within the magnetar model. According to this formulation, the energy injected by the spinning-down of a newly formed highly magnetic NS, commonly referred to as a magnetar, can produce the observed flattening in the LC. According to the formulation in \citep{ZhangMeszaros2001}, the luminosity emitted from the spinning-down can be written as:

\begin{equation}
    L_{sd}(t)=\frac{E_{NS}}{T(1+t/T)^2},
    \label{eq:magnetarluminosity}
\end{equation}

where $E_{NS}$ is the energy released by the NS and $T$ the time parameter $T \propto B^{-2} \omega^{-2}$ ($B$ being the magnetic field intensity and $\omega$ the rotational frequency). The luminosity $L_{sd}$ is proportional to the time $T^{-1}$, thus corroborating the luminosity-time relation \citep{Dainotti2013} consistent with the theoretical expectations. The energy injection stemming from the magnetar's spinning-down radiation in the forward shock fits accurately with approximately 90 $\%$ of the X-ray plateaus \citep{DallOsso2011,Bernardini2013,Rowlinson2013,Stratta2018} and the rotation period ($P$) values are consistent with the equilibrium values in the rotation period-magnetic field ($P-B$) relation $P_{eq} \propto B^{6/7} \dot{m}^{-1/2}$ (being $\dot{m}$ the accretion rate) expected from the millisecond pulsars but considering much higher values of $\dot{m}$ \citep{Pan2013}. In the lower panel of Figure \ref{fig:Stratta2018}, \citep{Stratta2018} show the $P-B$ diagram for a sample of 40 GRBs observed by Swift-XRT with regular plateau features (namely, with at least 5 data points at the beginning of the plateau phase and a plateau steepness less than $41^{\circ}$). In Figure \ref{fig:Stratta2018}, the fit is shown with line 2, while lines 1 and 3 mark the following mass accretion rates ($\dot{M}$) boundaries: $10^{-4}M_{\odot}s^{-1}<\dot{M}<0.1M_{\odot}s^{-1}$. Moreover, the explanation of the plateau emission through the magnetar spinning-down has been expanded in several relevant works \citep{Rowlinson2014,Rea2015}. It is crucial to note that the magnetar model is only one of the different ways to explain the plateau emission. However, additional work has been done to explain the plateau phase within the standard fireball model in multiwavelengths, namely, in X-rays \citep{Srinivasaragavan2020,Dainotti2021closure}, $\gamma$-rays \citep{Dainotti2023closure,Fraija2022sinch,Fraija2023closure}, optical \citep{Dainotti2022optical}, and radio \citep{Levine2023radio}. Indeed, the luminosity-time relation is also viable in optical \citep{Dainotti2020} and in radio \citep{Levine2022}.

Among all the observed LGRBs, the 210905A \citep{GRB210905A} represents an exceptional case. Characterized by a high $z$ value ($z=6.312$) and an isotropic energy release of $E_{iso}=1.27^{+0.20}_{-0.19}*10^{54}erg$, this GRB is among the most energetic in the Konus-Wind catalogue and is characterized by a bright X-ray, optical, and near-infrared afterglow. This GRB lies in the 68 $\%$ confidence region of the peak energy-isotropic energy relation, or Amati relation \citep{Amati2002} ($E_{peak}-E_{iso}$) while it also matches the predictions of the Yonetoku relation \citep{Yonetoku2004} ($E_{peak}-L_{iso}$) very accurately. Moreover, GRB 210905A is 1 $\sigma$ away from the expectations of the Ghirlanda relation \citep{Ghirlanda2004}. Thus, it proves to be compatible with a good number of correlations. Despite being a very bright event, it seems that the progenitor of this event is the same of low-$z$ GRBs rather than the Population III star progenitors. Furthermore, the jet's energy is too big to be due to a standard magnetar, and this implies that the central engine of this GRB could be a newly formed accreting BH.

\section{The observations of GRBs and the current and future missions}\label{sec:GRBmissions}
After discussing the current status of GRB astrophysics, it is essential to outline the future perspectives in GRB observations. The study of these transients deals with the open problem of selection biases that unbalance the distribution of discovered events in favour of low-$z$ events; see the discussion in \citep{DainottiPetrosian2013,Petrosian2015}. Therefore, using current and upcoming GRB missions to increase the sample size by including high-$z$ events is crucial. Here is an overview of ongoing and forthcoming missions.

\begin{figure}
    \centering
    \includegraphics[scale=0.395]{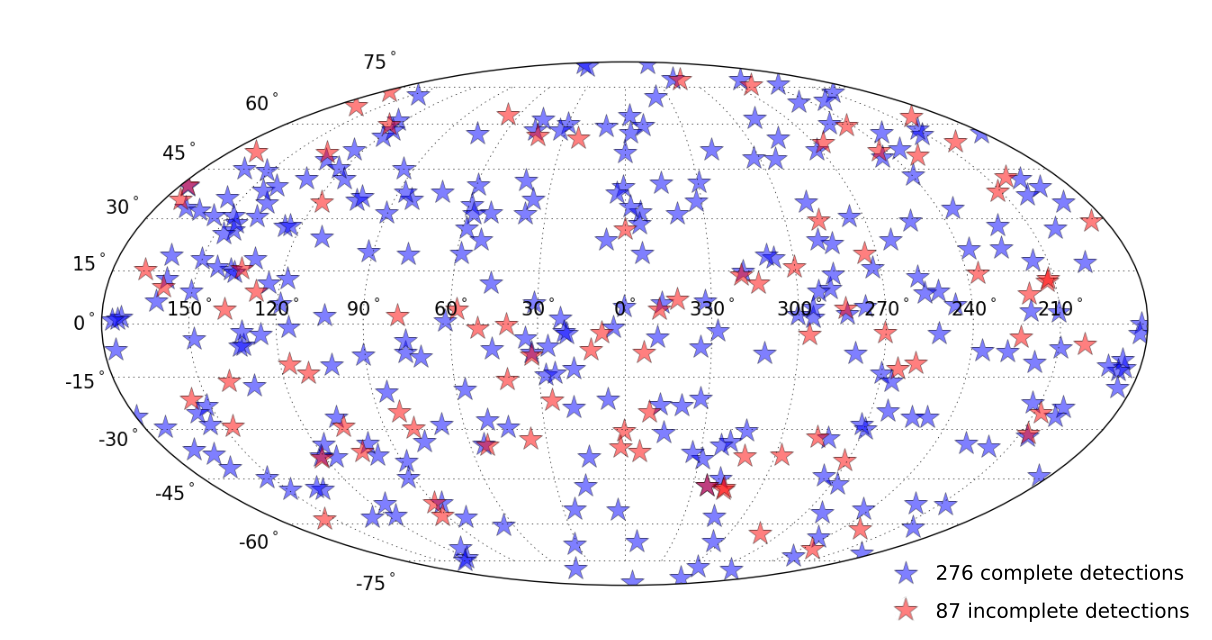}
    \includegraphics[scale=0.18]{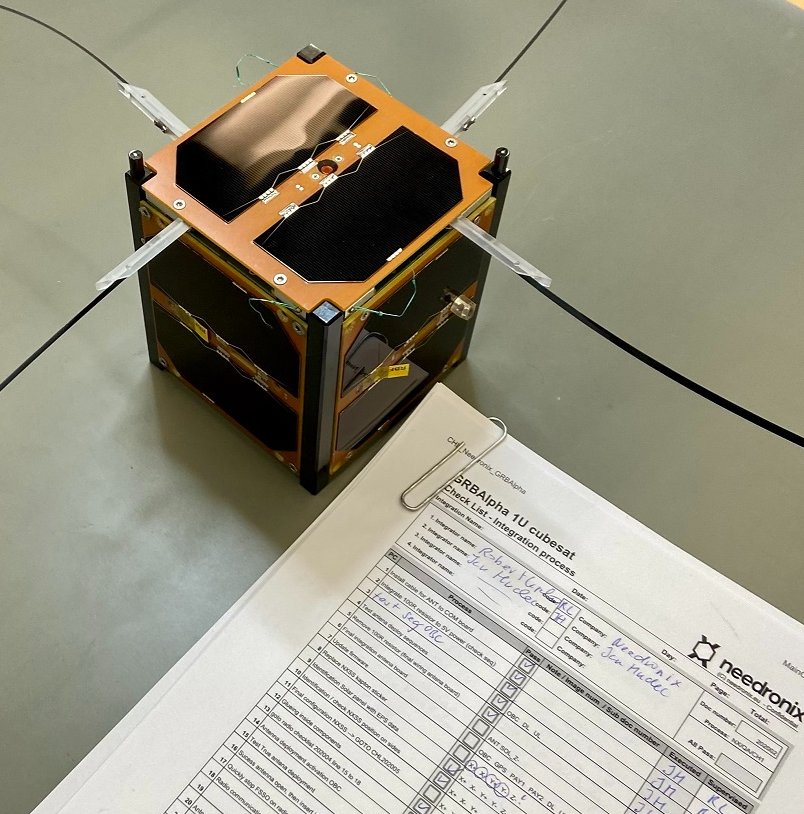}
    \includegraphics[scale=0.505]{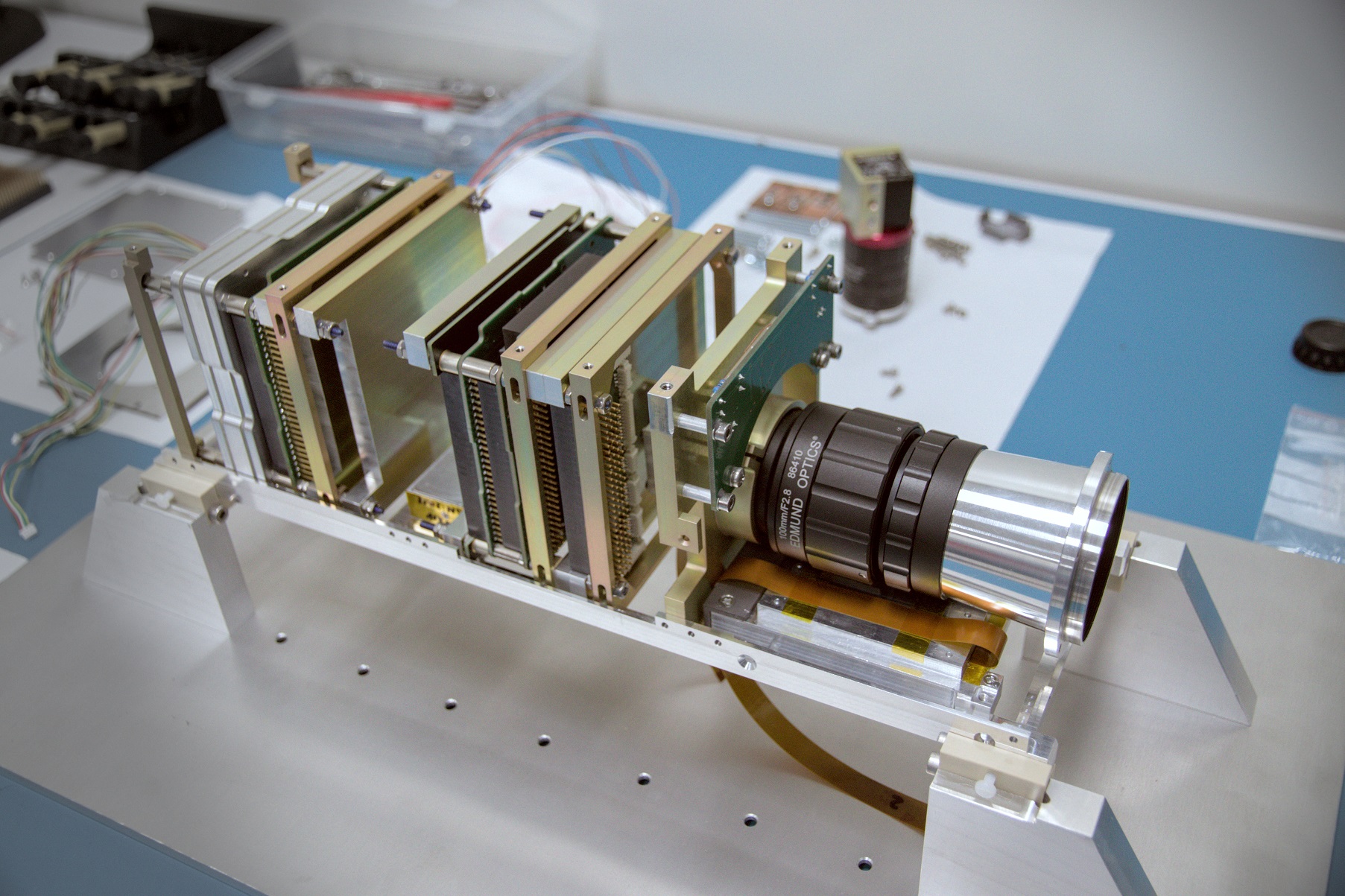}
    \includegraphics[scale=0.36]{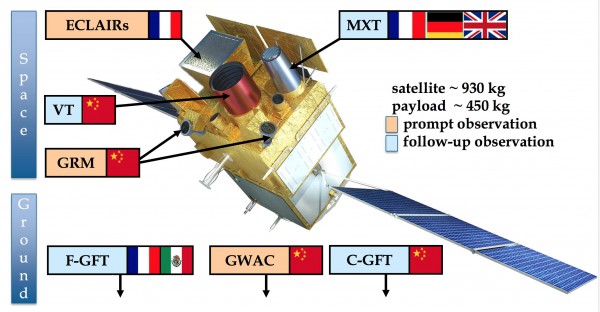}
    \includegraphics[scale=0.35]{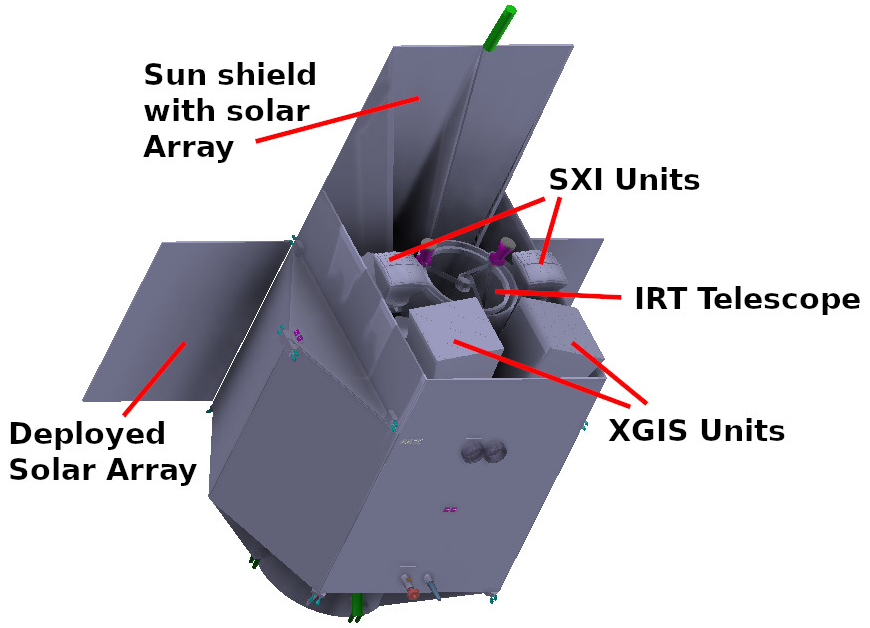}
    \caption{\textbf{Upper panel.} The map for the Second AGILE MCAL Gamma-Ray Burst Catalog. The 276 complete detections are reported with blue stars, while the incomplete ones are plotted with red stars. \textbf{Middle left panel.} The GRBAlpha satellite. \textbf{Middle right panel.} The VZLUSAT-2 satellite. \textbf{Lower left panel.} The SVOM satellite design. \textbf{Lower right panel.} The THESEUS mission design.}
    \label{fig:satellites1}
\end{figure}

The Astro‐Rivelatore Gamma a Immagini Leggero (AGILE) $\gamma$-ray satellite, an initiative of the Italian Space Agency (ASI) launched in 2007, has been operational in space for 16 years. Over this duration, AGILE has observed and recorded data from more than 500 GRBs, as documented in the AGILE Mini-Calorimeter (MCAL) Second GRB catalogue release \citep{Ursi2022b}.
Figure \ref{fig:satellites1}'s upper panel presents a visualization of 363 GRBs from the Second AGILE MCAL Gamma-Ray Burst Catalog. Among these, 276 GRBs represent complete detections, namely, the cases where the onboard triggered data acquisition encompassed the entire duration of the GRB. However, 87 of these events constitute incomplete bursts, where only partial or fragmented detection was obtained.
AGILE's observations have extended beyond routine GRBs, encompassing distinctive events such as the luminous 190114C \citep{Ursi2020}, the New Year's Burst 220101A \citep{Ursi2022c}, and the Brightest Of All Times (BOAT) 221009A \citep{Tavani2023}. Furthermore, AGILE has significantly contributed to identifying EM counterparts for events catalogued in the GWTC-1 catalogue \citep{Ursi2022a}. It is anticipated to further bolster research efforts during the O4 run of LVK, specifically concerning these events.
The contributions of AGILE extend beyond GRB studies, encompassing fields like FRBs with observations of SGR1935+2154 \citep{Tavani2021}, terrestrial gamma-ray flashes (TGFs) as documented in the Third AGILE TGF Catalog, and solar physics through The First AGILE Catalog of Solar Flares \citep{Ursi2023ApJS}. These diverse contributions underscore AGILE's multifaceted role in advancing our understanding across various astrophysical domains.

An intriguing perspective emerges from the GRBAlpha and VZLUSAT-2 CubeSat nanosatellites: the former, a one unit (1-U) CubeSatellite equipped with a gamma-ray detector, has observed 34 GRBs, 2 soft gamma repeaters (SGRs), and 16 solar flares since its launch in 2021. Conversely, the latter, a 3-U satellite housing two detectors launched in January 2022, recorded 20 GRBs, 5 SGRs, and 20 solar flares. Notably, the GRBAlpha satellite successfully observed the event associated with the BOAT 221009A, registering a flux of $22000\,count/s$ within the $80-950\,{\rm keV}$ energy band \citep{Ripa2023}, while the VZLUSAT-2 detected GRB 230307A linked to a KN event \citep{GCN33424}. These pioneering missions aim to evaluate the potential for localizing GRBs using a forthcoming fleet of nanosatellites. In Figure \ref{fig:satellites1}, the GRBAlpha and VZLUSAT-2 satellites are depicted in the left and right middle panels, respectively.

Furthermore, the forthcoming SVOM mission \citep{Bernardini2021}, a collaborative Franco-Chinese initiative set for launch in March 2024, represents another substantial advancement in this domain. SVOM will deploy four primary instruments: ECLAIRs for detecting energies ranging from $4$ to $250\,{\rm keV}$, the Microchannel X-ray Telescope (MXT) targeting the $0.2$ to $10\,keV$ range, the Gamma Ray Burst Monitor (GRM) for spectral measurements between $15\,{\rm keV}$ to $5000\,{\rm keV}$, and the Visible Telescope (VT) aimed at identifying burst counterparts within the visible spectrum. Weighing 930 kg with a 450 kg payload, SVOM will orbit at a low earth inclination of $30^{\circ}$, maintaining an altitude of 625 km and an orbital period of 96 minutes. The lower left panel of Figure \ref{fig:satellites1} represents the satellite's design configuration. This upcoming mission heralds a significant leap in capabilities to investigate and analyze celestial transient events, particularly within the GRB domain.

In the early 2030, if the mission is approved, significant advancements in the astrophysical study of GRBs are anticipated from the Transient High-Energy Sky and Early Universe Surveyor (THESEUS) \citep{MereghettiBalman}. This satellite mission encompasses an X-ray telescope operating within the $0.3-5\,{\rm keV}$ range utilizing lobster-eye focusing optics, a gamma-ray spectrometer spanning the $2-150\,{\rm keV}$ band, and a near-infrared telescope measuring 0.7 meters. THESEUS was recently selected for the European Space Agency's (ESA) phase-0 study in the M7 selection process, which is projected to extend until 2025. THESEUS is currently concurring with other two mission concepts selected in November 2023 for a 3-year Phase and, if approved, its launch will be set for 2037. The mission's design layout is depicted in the lower right panel of Figure \ref{fig:satellites1}.

\begin{figure}
    \centering
    \includegraphics[scale=0.07]{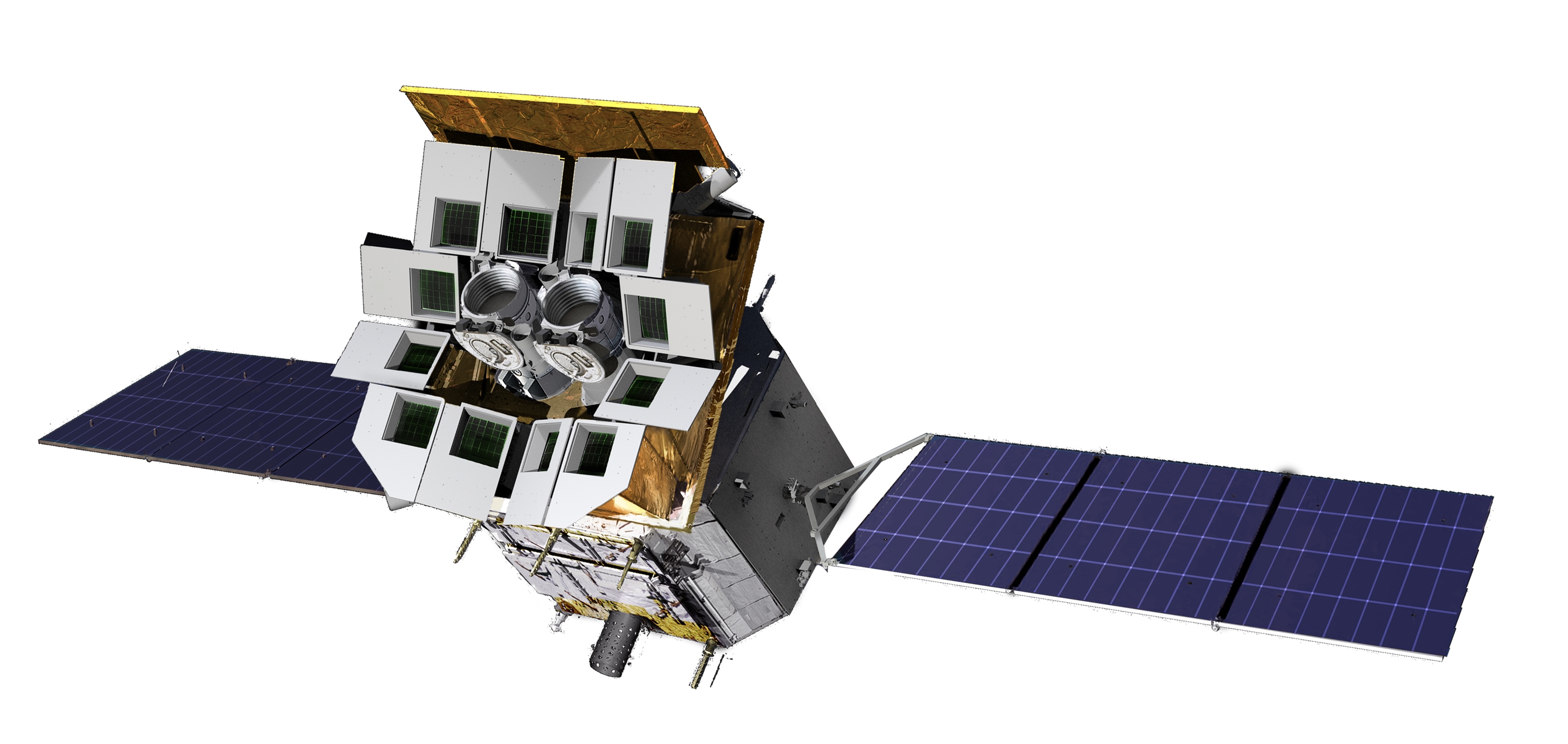}
    \includegraphics[scale=0.33]{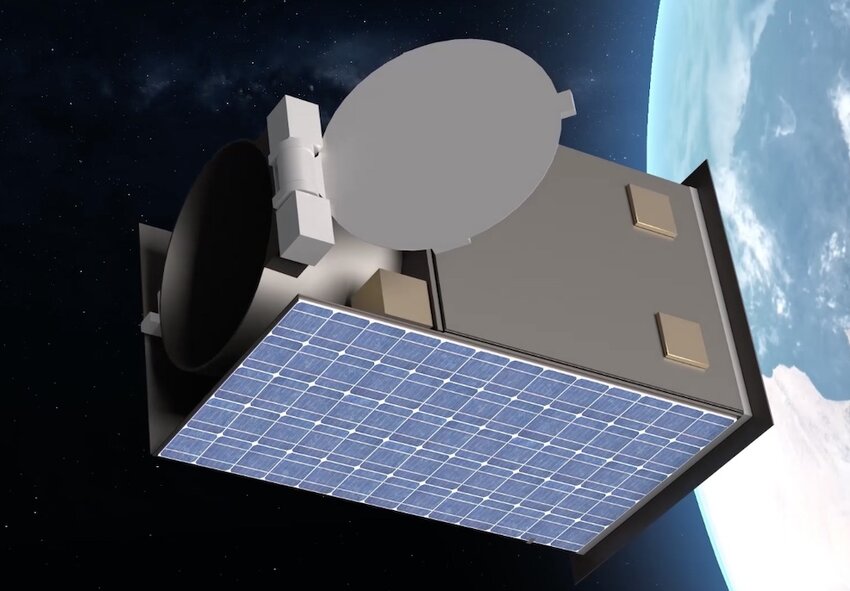}
    \includegraphics[scale=0.57]{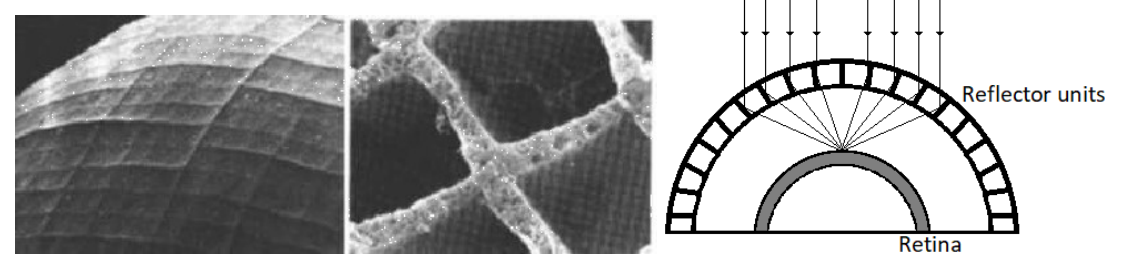}
    \caption{\textbf{Upper left panel.} The Einstein Probe mission concept. \textbf{Upper right panel.} The QUVIT satellite design. \textbf{Lower panel.} An electron microscope image showing the lobster's eye surface, together with a diagram of its working.}
    \label{fig:satellites2}
\end{figure}

Another noteworthy mission on the horizon is the Einstein Probe \citep{Boller}, a collaborative initiative involving the Chinese Academy of Sciences (CAS), ESA, and the Max Planck Institute for Extraterrestrial Physics (MPE). Illustrated in the upper left panel of Figure \ref{fig:satellites2}, this satellite, slated for launch by the end of 2023, boasts an energy resolution range of 0.5-5 keV and the capacity for follow-up observations spanning 0.3-10 keV. The Einstein Probe is poised to make substantial contributions across a spectrum of astrophysical phenomena, encompassing GRBs, SN shock breakout events, magnetar signals, Fast Radio Bursts (FRBs), X-ray flares from stars, thermal nuclear bursts, X-ray binaries (XRBs), IMBH accretion signals, AGNs, quasi-periodic X-ray eruptions, and X-ray flares associated with neutrino detections. This mission holds significant promise in advancing our understanding of many cosmic phenomena within the X-ray domain.

It is noteworthy to highlight recent advancements in X-ray observation technologies, particularly in developing innovative designs such as the Lobster Eye (LE) X-ray Optics \citep{Hudec2022}, a configuration that is aimed to observe the X-ray afterglows of GRBs. This optical system, inspired by the structure of lobsters' eyes, is constructed from millions of square reflection canals, as illustrated in the lower panel of Figure \ref{fig:satellites2}. Employing Micro Pore Optics (MPOs), the LE X-ray Optics consists of glass optics, typically ranging in 1-3 mm thickness, featuring regularly arranged square pores set in a spherical configuration and coated to reflect X-rays effectively. Each MPO unit typically measures 40x40 mm, with pores spanning 20-40$\mu m$ in width and exhibiting an open fraction of 60 $\%$ or higher. This configuration enables a larger Field of View (FoV) compared to conventional Wolter telescope designs while maintaining a significantly lower mass. Such a configuration allows an optimal detections of phenomena such as AGNs, galactic bulge sources, XRBs, SGRs, GRBs, X-ray Novae, and X-ray flares generated from stars.

Regarding the observations of new KNe, this task relies on missions like the Quick Ultra-Violet Kilonova surveyor (QUVIK) \citep{Werner2022}, represented in the upper right panel of Figure \ref{fig:satellites2}. QUVIK, a near-ultraviolet space telescope mounted on a 70 kg microsatellite, is tailored to capture the evolution of KNe within the near-ultraviolet ($260-360\,nm$) and far-ultraviolet ($140-190\,nm$) spectra, equipped with a 33-centimeter primary mirror configured in a modified Cassegrain scheme. This mission aims to survey merging events generating KNe within a range of 200 Mpc from Earth, anticipating an observation rate of several tens of events annually. QUVIK is undergoing a phase B1 study, with a scheduled launch anticipated between late 2027 and early 2028. This mission stands poised to augment our understanding of KNe astronomical events significantly.

\section{The applicability of the magnetar model}\label{sec:magnetar}
The magnetar model offers an explanatory framework for the phenomenon of GRB plateau emission and provides insights into the observation of X-ray bursts. These transients occurr with durations in the range $millisec-sec$ and luminosities of about $10^{36}-10^{43}\,erg\,s^{-1}$ and can be explained in the context of Alfvén waves, namely, plasma waves where there is a propagation of the ion oscillations. In \citep{YuanBeloborodov}, the authors perform 3D simulations in the force-free electrodynamics regime for a localized Alfvén wave packet launched by a magnetar quake into the magnetosphere. According to these simulations, the model can explain the observed spikes coincident with the FRB-like detection of SGR 1935+2154, which has been generated by a galactic magnetar. Their simulations highlighted that if the Alfvén packet, while propagating to a radius $R$, possesses total energy surpassing the magnetospheric energy $B^{2}R^{3}/8\pi$, the wave can undergo non-linear behaviour and be expelled from the magnetosphere. This expulsion leads to the displacement of magnetospheric field lines and creates a reconnection zone for these lines. Subsequent reconnection events in this zone facilitate plasma energization, producing observable X-ray emissions from radiative shocks. This causal link between the Alfvén wave dynamics and the observed X-ray bursts offers valuable insights into the mechanisms at play within magnetar-driven astrophysical events.

\section{The accretion mechanism as the engine of AGNs}\label{sec:accretion}
In the preceding Section, the discussion highlighted the magnetar model's capability to elucidate various GRB observations but there are exceptions, represented for example by the case of GRB 210905A, which can be explained within the fallback accretion onto a newly formed BH. This Section begins with introducing a comprehensive formulation to describe the accretion mechanism.
\citep{Bisnovatyi-Kogan} derive an equation for the time lag between the X-ray and optical maximum emission in a cosmic source based on the disk-accretion mechanism. The time lag $\tau$ can be expressed as

\begin{equation}
    \tau=6.9 \frac{m^{2/3}\dot{m}^{1/15}}{\alpha^{4/5}(T_4)^{28/15}}\,days,
    \label{eq:timelagaccretion}
\end{equation}

where $m$ is the mass of the object that accretes (in $M_\odot$), $\dot{m}$ is the time-derivative of $m$ (expressed in units of $10^{-8}M_\odot/year$, $T_4$ is the maximum temperature in the optics (in units of $10^4\,K$), and $\alpha$ is the viscosity of the accreting disk. This formulation boasts broad applicability across cosmic entities governed by disk accretion mechanisms. The observed emissions align with this model when the disruption of a giant-phase star induces entry into regions of intense tidal forces, allowing material with low angular momentum to accrete onto the supermassive black hole (SMBH). The variability observed in many AGNs, where optical-ultraviolet signals lag the X-ray emission, can be predicted when an X-ray flash in the centre of the AGN is followed by re-radiation of the accretion disk. This theoretical framework offers valuable insights into the temporal dynamics of emissions observed in diverse cosmic phenomena regulated by accretion processes.

The SMBH in the core of massive galaxies critically affects their physical manifestation: when the surrounding material is captured by the BH, an accretion disk is generated and sometimes the BH ejects plasma as an outflow or a collimated jet \citep{Panessa2019}. The resulting radio emission in the AGNs spans over a factor of $10^5$ between the faintest and the brightest ones. The radio-quiet (RQ) AGNs are usually 1000 times fainter than the radio-loud (RL) ones, and the former allows the investigation of different mechanisms for the radio emission when a luminous jet is not hiding the other signals in the AGN. In the RQ objects, four possible mechanisms can produce the radio emission in the observed amounts: 

\begin{itemize}
    \item Ejected materials from the accreting SMBH in the form of collimated and energetic jets;
    \item AGN wind shocks;
    \item Star formation that produces both thermal and non-thermal radio emission;
    \item The hot corona around the SMBH, suggested by the correlation between the radio and X-ray luminosities ($L_R$ and $L_X$, respectively) of the AGN, namely the Neupert effect (where $L_R$ = $dL_X/dt$).
\end{itemize}

\begin{figure}
    \centering
    \includegraphics[scale=0.45]{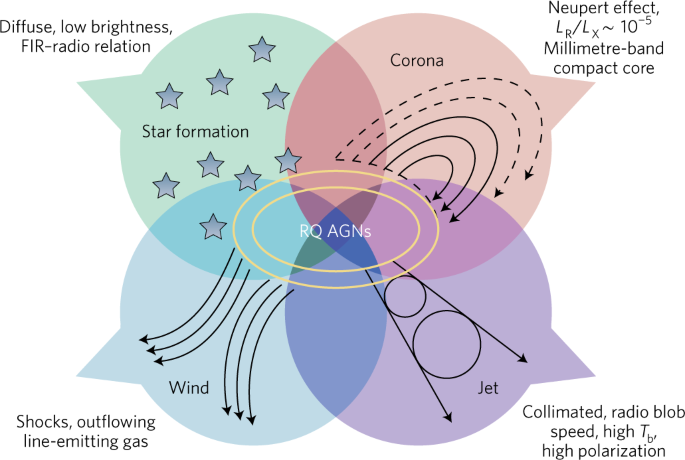}
    \includegraphics[scale=1.25]{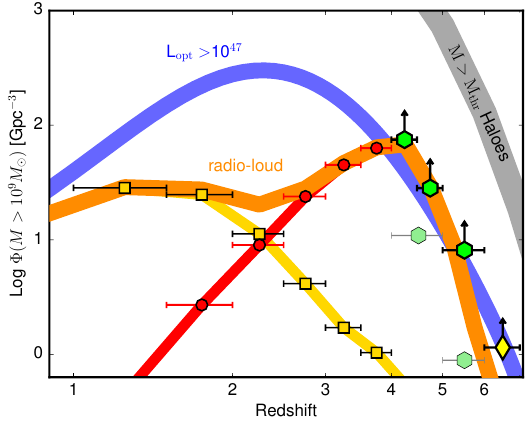}
    \caption{\textbf{Upper panel.} The four mechanisms for explaining the radio emission in the RQ AGN cases. \textbf{Lower panel.} The comoving number density for $M \geq 10^{9}M_\odot$ SMBHs observed in the jetted AGNs and all the AGNs.}
    \label{fig:AGNs}
\end{figure}

The four mechanisms presented above are depicted in the upper panel of Figure \ref{fig:AGNs}. The RQ AGNs are variable in the radio emission, and this opens new observational classes: the changing-look AGNs, the tidal disruption events (TDE) AGNs, and the flaring RQ AGNs. 
\citep{Sbarrato2022} estimate the comoving number density of highly accreting SMBH with extreme masses $M \geq 10^{9}M_\odot$ in both the jetted and non-jetted AGN cases. In the lower panel of Figure \ref{fig:AGNs}, the SMBH hosted in the jetted AGNs are in the orange line, while the ones hosted in all the AGNs are in the blue line.
To draw the non-jetted curve, the bolometric quasar luminosity function (QLF) in the range $0<z<7$ \citep{Shen2020} is taken as a reference. The RL sample is built up of Fermi-LAT observations (yellow line and squares) and Swift-BAT (red line and circles). The bright green hexagons are the lower limits to the $z>4$ comoving number from the Sloan Digital Sky Survey and the Faint Images of the Radio Sky at Twenty-centimeters Survey combined (SDSS+FIRST). The pale hexagons show the status before the beginning of a systematic search for blazars at high-$z$. The yellow diamond is the lower limit given by the blazar PSO J0309+27 \citep{Belladitta2020}.
According to these observations, the jetted AGNs seem more numerous than the non-jetted AGNs at $z>4$: this implies that the jets must have a prominent role in the rapid creation of the first SMBH in the universe. 
The Imaging X-ray Polarimetry Explorer (IXPE) has enormously contributed to the observations of AGNs, pulsars, and magnetars. IXPE observed the RQ AGNs NGC 4151, Circinus Galaxy, MCG-5-23-16, and IC 4329A as NGC 4151 and Circinus Galaxy, respectively, as a Seyfert Type 1 and 2 galaxy. For what it concerns, the strong pulsars, Her X-1, Cen X-3, and X Persei have been observed by IXPE, as well as the weak pulsars 4U1626-67 \citep{Marshall}, GRO J1008-57, and GX 301-2. Furthermore, IXPE has also targeted weak-field NSs, such as the Cyg X-2.

\section{The advent of neutrino astronomy in the multimessenger era}\label{sec:neutrino}
In the field of high energy astrophysics, we have also the neutrino astronomy, offering valuable insights into extragalactic phenomena. As an exciting example, it is worth mentioning the neutrino observations of the galaxy NGC 1068, a Seyfert II galaxy, with a series of excess neutrinos observed from the IceCube telescope \citep{Abbasi2022} that are characterized by a significance of $4.2\sigma$. The remaining events detected by IceCube exhibit a significance level of $\sim 3\sigma$, with most associations traced back to blazars. These fall into two primary categories: low-energy-peaked objects (LBLs) and intermediate-high-energy-peaked objects (IHBLs), with a preference for these latter as neutrino sources through the proton acceleration mechanism.

In \citep{Gasparri2022}, an investigation encompassed 16 IceCube neutrino events spanning September 2018 to March 2020. This research aimed to identify potential counterparts within the publicly accessible AGILE archive spanning the entire mission duration. Out of the total 16 LCs identified, 8 of them show a significant $\gamma$-ray detection: $2/3$ Extremely High Energy (EHE) neutrinos (IC-180908A and IC-190503A), $3/6$ High Energy Starting Event (HESE) neutrinos (IC-190104A, IC-190221A, and IC-190504A), $3/7$ Gold neutrinos (IC-190619A, IC-190922A, and IC-191001A) with great significance. Furthermore, 2 out of 16 are associated with the 2AGL Catalog sources.
This research suggests intriguing correlations between neutrino events observed by IceCube and $\gamma$-ray detections in AGILE, shedding light on potential extragalactic neutrino sources.

\section{The young universe between observations with space telescopes and cosmology}\label{sec:cosmologyobs}
The current Section shifts the focus to the observational and theoretical aspects of the large-scale universe. Recent observations from the James Webb Space Telescope (JWST) and the Hubble Space Telescope (HST) suggest that massive primordial black holes (PBH) could represent the seed for the galaxies and quasars in the very young universe, as discussed in \citep{Dolgov2023} and previously proposed in \citep{DolgovSilk1993}. Indeed, the log-normal mass spectrum for PBH obtained from this proposal is in excellent agreement with the LVK observations.

Another compelling cosmology observation involves the intricate relationship between cosmic rays and DM. In standard cosmological models, stable supersymmetric relics, regarded as potential DM particle candidates, are believed to have masses below $1\,{\rm TeV}$. However, this presumption contradicts constraints established by the Large Hadron Collider (LHC) concerning low-energy Supersymmetry (SUSY). The study conducted in \citep{Arbuzova2022} extends the possible mass range for DM particles to encompass superheavy values. \citep{Arbuzova2022} propose that these particles could exist within a mass spectrum spanning from $10^{6}-10^{13}\, {\rm GeV}$, depending on the prevalent decay mode of the scalaron. The potential decay of these particles, possibly through virtual Black Holes (BH) formation, could generate superheavy relics that are stable enough to behave like DM. Moreover, the decay of these relics might explain the observed properties of cosmic rays.

Advancing our knowledge of cosmological structures involves a comprehensive investigation into the filamentary architecture constituting the Cosmic Web and the endeavour to ascertain the whereabouts of missing baryonic matter. To this end, it is essential to study the observations of the Sunyaev Zel'dovich effect. This effect represents the boost of Cosmic Microwave Background (CMB) photons that receive energy from the electrons of hot gases diffused in the universe through the Inverse Compton scattering. The 90 GHz MISTRAL instrument will enhance the study of this effect: this camera was commissioned in 2022 on the 64-meter ground-based Sardinia Telescope \citep{Battistelli2022}. The MISTRAL instrument has an angular resolution of 12 arcsec, way more precise than the angular scale of Planck (10 arcmin). The instrument is visible in the lower panel of Figure \ref{fig:cosmology}.

\section{An alternative cosmological paradigm}\label{sec:alternativecosmology}
\begin{figure}
    \centering
    \includegraphics[scale=0.38]{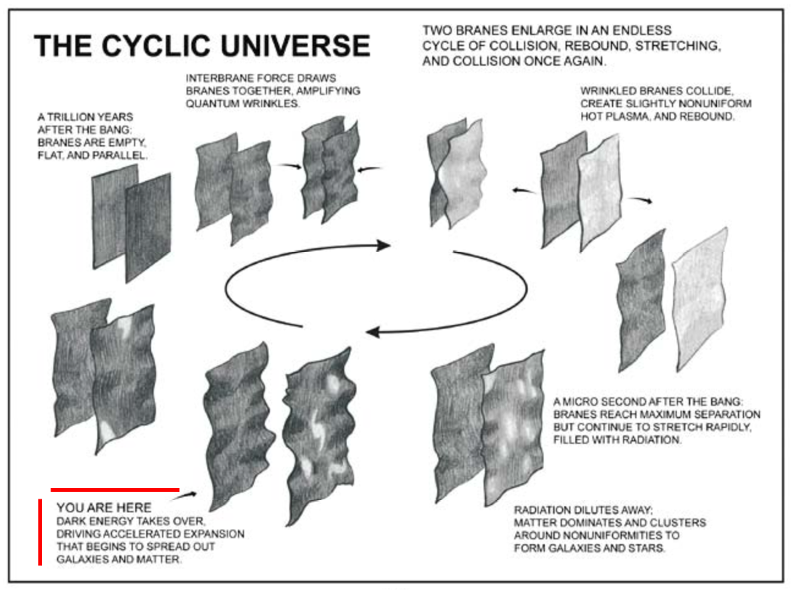}
    \includegraphics[scale=0.37]{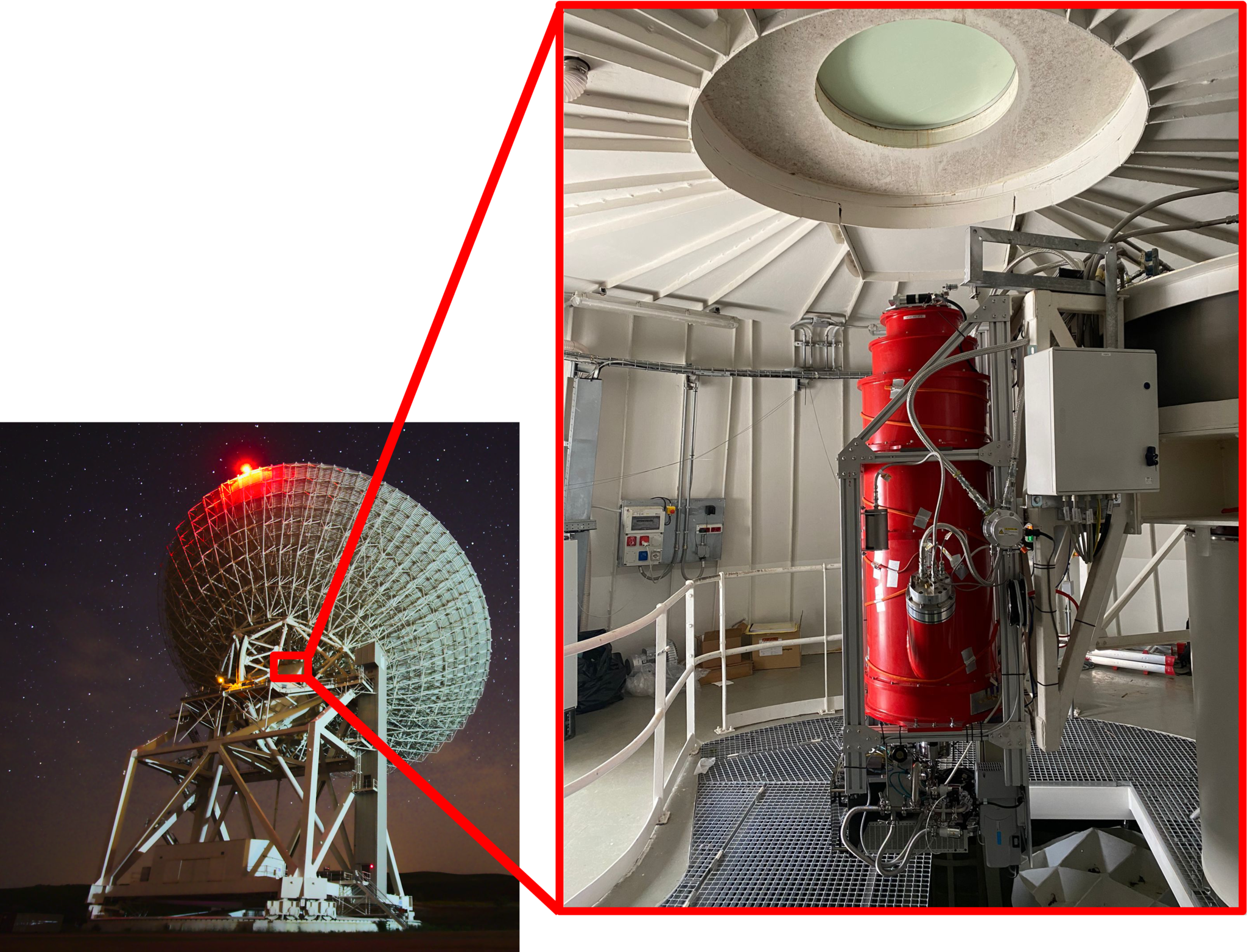}
    \caption{\textbf{Upper panel.} The summary of the Membrane-Universe theory. \textbf{Lower panel.} The MISTRAL instrument shown together with the 64-meters Sardinia Telescope.}
    \label{fig:cosmology}
\end{figure}

Theoretical frameworks designed to elucidate the origin and expansion of the universe are mandated to respect the experimental observation of the inflationary phase \citep{Giovannelli2016}. This phenomenon can be proved by measuring CMB polarization. Nevertheless, some alternative scenarios, like the Membrane-Universes theory, do not need the inclusion of an inflationary phase \citep{Giovannelli2019}. In this framework, the universes clash endlessly and provide a cyclic model for the observable universe. The model is advantageous since it can overcome the necessity of defining the origin of time and the setting of initial conditions. Furthermore, the cyclic-universe scenario shows how the galaxies and the large scale structure present in one cycle are created only by the quantum fluctuations in the preceding cycle, being independent on the earlier cycles and without affecting any structure in the later cycles \citep{Erickson2007}. The upper panel of Figure \ref{fig:cosmology} summarises the Membrane-Universes theory.

\section{Back to the local scales}\label{sec:local}
We now pass from the farway universe to the local scale. This Section highlights contributions to examining local-scale objects observable within our Galaxy. 

\subsection{Core Collapse SNe remnants}
Core-collapse supernova Remnants (SNRs) are of particular interest, owing to their properties that offer insights into SN engines, progenitor star attributes, interactions within the circumstellar medium (CSM), and final stages of stellar evolution. \citep{Orlando2023} propose an integrated methodology that coherently leads from the progenitor massive star to the SNe and SNR observations. Employing 3D hydrodynamic and magnetohydrodynamic simulations, this method models steps from the SN explosion to SNR formation, incorporating data from observations about the progenitor star and CSM. This comprehensive approach integrates observed dynamics, energetics, and spectral properties of SNe and SNRs into a single model.

\subsection{Galactic Wolf-Rayet binary stars}
In the investigation of binary Wolf-Rayet (WR) stars, \citep{MiyamotoIshida} analyze a series of X-ray Multi-Mirror Mission Reflection Grating Spectrometer (XMM-Newton RGS) data of the XRB WR140. The study determines the geometry of the shock cone from the ram-pressure balance between the stellar winds at each phase. It measures the line-of-sight velocity ($v_{los}$) together with the velocity dispersion ($\sigma_{los}$). The authors identify the Ne-line emission by comparing the theoretical value of $v_{los}$ with the observed one. Employing this approach across multiple emission lines facilitates the derivation of the plasma's temperature and density profiles along the shock cone, thereby providing comprehensive insights into its characteristics.

\subsection{The globular clusters}
\begin{figure}
    \centering
    \includegraphics[scale=0.25]{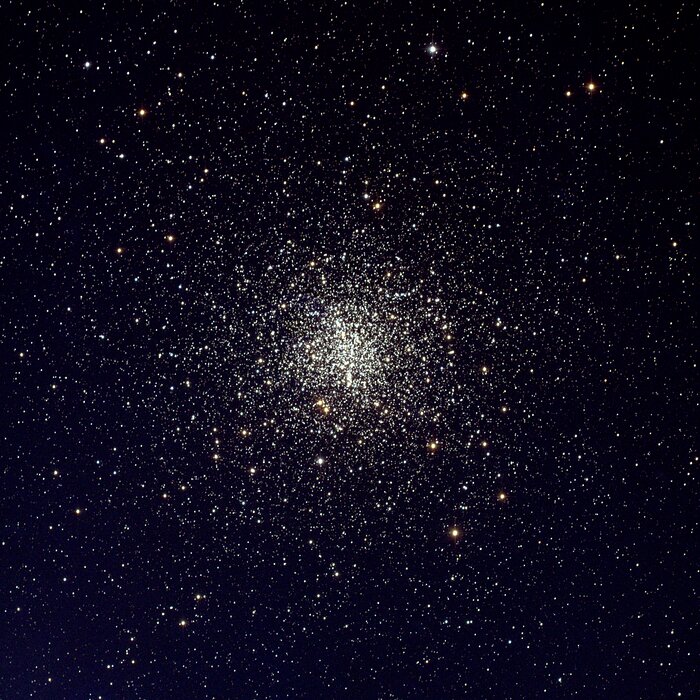}
    \caption{An image of the globular cluster NGC 6121 (Credits: NOIRLab/NSF/AURA).}
    \label{fig:local}
\end{figure}

The globular clusters represent another exciting class of objects in our Galaxy. \citep{Merafina} aim to extract information from the proper motions of the stars to obtain clues on the effects of the tidal forces induced by the host Galaxy on the globular cluster. The object of study is the globular cluster NGC 6121 (Figure \ref{fig:local}) through the Global Astrometric Interferometer for Astrophysics (Gaia) satellite measurements: after cleaning the sample, with the exclusion of stars affected by negative parallaxes and huge relative uncertainties, the initial database of 28000 stars is reduced to 8000 stars. The results obtained in this analysis agree with the theoretical predictions. Nevertheless, unresolved issues persist, including the potential incompleteness of the utilized data set, reliance on a singular mass function, and the assumption of a constant gravitational potential within a fixed shell, indicating avenues for further investigation and refinement.

\subsection{The detection of exoplanets from the Kepler mission}
To conclude this Section, we move to the study of exoplanets and their discovery by the Kepler mission \citep{Kepler}. As of June 30, 2022, the Kepler mission has collected 8793 candidate exoplanets, 5054 of these being confirmed and 3789 being in planetary systems like our Solar System. It is interesting to highlight that, among all the candidate exoplanets, 941 have a radius below the threshold of 1.25 Earth radius, thus implying the existence of several Earth-like planets in the Milky Way and, thus, the possible existence of extraterrestrial life in our Galaxy.

\section{Summary and conclusions}\label{sec:conclusions}
The compilation of contributions in this collection of Research Highlights presents an extensive overview of current astrophysical and cosmological observations. A critical overarching insight derived from these studies underscores the significance of multimessenger astronomy. Initiated by GW observations, the pivotal connection with additional sources of information (EM, neutrino) emerges as indispensable for further constraining the physics governing observed phenomena's progenitors. GRBs assume a pivotal role in this context due to their multiwavelength observations and their association with events like GRB 170817A, correlated with GW170817 and AT 2017gfo triggers.
The exploration of transient and high-energy astrophysical phenomena accentuates the critical need for robust theoretical frameworks. Investigations into magnetar spin-down mechanisms and accretion scenarios are pivotal focal points in this domain.
Objects examined thus far originate from cosmological realms, crucial for probing the early universe's nature and structure. The utilization of space telescopes allows for the indirect observations of the very first black holes and how they formed. In tandem with direct observations, alternative cosmological models such as the Membrane-Universe theory and theoretical propositions elucidating the nature of DM play a crucial role in unravelling the constituents of our universe.
Shifting the focus to local scales, recent discoveries within SNRs, XRBs, globular clusters, and exoplanets encapsulate significant strides in unravelling the mysteries that pertain to the local-scale universe. These findings hold substantial importance in broadening our understanding of local-scale cosmic phenomena.


\begin{thebibliography}{99}

\bibitem{GW170817}
B. P. Abbott et al. 2017, \emph{GW170817: Observation of Gravitational Waves from a Binary Neutron Star Inspiral}, Phys. Rev. Lett., 119 161101

\bibitem{Rosinska2023}
D. Rosinska 2023, \emph{Gravitational Waves Astronomy:
present and future}, Multifrequency Behaviour of High Energy Cosmic Sources - XIV, 12-17 June 2023

\bibitem{Poggiani2023}
R. Poggiani 2023, \emph{The LIGO-Virgo O3 Run and the Multi-Messenger Investigations of Compact Binary Mergers}, ANNALEN DER PHYSIK, 2200215, https://doi.org/10.1002/andp.202200215

\bibitem{Szczepanczyk2023}
M. J. Szczepańczyk et al. 2023, \emph{An Optically Targeted Search for Gravitational Waves emitted by Core-Collapse Supernovae during the Third Observing Run of Advanced LIGO and Advanced Virgo}, eprint arXiv:2305.16146

\bibitem{GRB170817A}
B. P. Abbott et al. 2017, \emph{Gravitational Waves and Gamma-Rays from a Binary Neutron Star Merger: GW170817 and GRB 170817A}, ApJL, 848 2 L13 

\bibitem{AT2017gfo}
S. Valenti et al. 2017, \emph{The Discovery of the Electromagnetic Counterpart of GW170817: Kilonova AT 2017gfo/DLT17ck}, ApJL, 848 2 L24

\bibitem{PatricelliBernardini}
B. Patricelli et al. 2022, \emph{Prospects for multi-messenger detection of binary neutron star
mergers in the fourth LIGO-Virgo-KAGRA observing run
}, MNRAS, 513 3 4159–4168

\bibitem{Rossi2020}
A. Rossi et al. 2020, \emph{A comparison between short GRB afterglows and kilonova AT2017gfo: shedding light on kilonovae properties}, MNRAS, 493 3 3379-3397

\bibitem{Vergani}
S. D. Vergani et al. 2017, \emph{The chemical enrichment of long gamma-ray bursts nurseries up to $z = 2$}, A\&A, 599 A120

\bibitem{GhirlandaSalvaterra}
G. Ghirlanda \& R. Salvaterra 2022, \emph{The Cosmic History of Long Gamma Ray Bursts}, ApJ, 932 10

\bibitem{Cucchiara2011}
A. Cucchiara et al. 2011, \emph{A photometric redshift of $z \sim 9.4$ for GRB 090429B}, ApJ, 736 7

\bibitem{Dainotti2008}
M. G. Dainotti et al. 2008, \emph{A time–luminosity correlation for $\epsilon$-ray bursts in the X-rays}, MNRAS, 391 1 79-83

\bibitem{Dainotti2011}
M. G. Dainotti et al. 2011, \emph{Towards a standard gamma-ray burst: tight correlations between the prompt and the afterglow plateau phase emission}, ApJ, 730 135

\bibitem{Dainotti2013}
M. G. Dainotti et al. 2013, \emph{Determination of the intrinsic Luminosity Time Correlation in the X-ray Afterglows of GRBs}, ApJ, 774 157

\bibitem{Dainotti2015}
M. G. Dainotti et al. 2015, \emph{Luminosity-time and luminosity-luminosity correlations for GRB prompt and afterglow plateau emissions}, MNRAS, 4514 3898-3908

\bibitem{Dainotti2016}
M. G. Dainotti et al. 2016, \emph{A Fundamental Plane for Long Gamma-Ray Bursts with X-Ray Plateaus}, ApJL, 825 L20

\bibitem{Dainotti2017}
M. G. Dainotti et al. 2017, \emph{A study of gamma ray bursts with afterglow plateau phases associated with supernovae}, A\&A, 600 A98 11

\bibitem{Dainotti2022ApJS}
M. G. Dainotti et al. 2022, \emph{The Optical Two- and Three-dimensional Fundamental Plane Correlations for Nearly 180 Gamma-Ray Burst Afterglows with Swift/UVOT, RATIR, and the Subaru Telescope}, ApJS, 261 25

\bibitem{Dainotti2020}
M. G. Dainotti et al. 2020, \emph{The Optical Luminosity–Time Correlation for More than 100 Gamma-Ray Burst Afterglows}, ApJL, 905 L26

\bibitem{Dainotti2022MNRAS}
M. G. Dainotti et al. 2022, \emph{Optical and X-ray GRB Fundamental Planes as cosmological distance indicators}, MNRAS, 514 2 1828-1856

\bibitem{Dainotti2022PASJ}
M. G. Dainotti et al. 2022, \emph{Gamma-ray bursts, supernovae Ia, and baryon acoustic oscillations: A binned cosmological analysis}, PASJ, 74 5 1095-1113 


\bibitem{Bargiacchi2023a}
G. Bargiacchi et al. 2023, \emph{Gamma-ray bursts, quasars, baryonic acoustic oscillations, and supernovae Ia: new statistical insights and cosmological constraints}, MNRAS, 521 3 3909-3924

\bibitem{Bargiacchi2023b}
G. Bargiacchi et al. 2023, \emph{Tensions with the flat $\Lambda$CDM model from high-redshift cosmography}, MNRAS, 525 2 3104-3116

\bibitem{Cao2022a}
S. Cao et al. 2022, \emph{Standardizing Platinum Dainotti-correlated gamma-ray bursts, and using them with standardized Amati-correlated gamma-ray bursts to constrain cosmological model parameters}, MNRAS, 512 1 439–454

\bibitem{Cao2022b}
S. Cao et al. 2022, \emph{Gamma-ray burst data strongly favour the three-parameter fundamental plane (Dainotti) correlation over the two-parameter one}, MNRAS, 516 1 1386-1405

\bibitem{DainottiBargiacchi2023}
M. G. Dainotti et al. 2023, \emph{Reducing the uncertainty on the Hubble constant up to 35\% with an improved statistical analysis: different best-fit likelihoods for Supernovae Ia, Baryon Acoustic Oscillations, Quasars, and Gamma-Ray Bursts}, ApJ, 951 63

\bibitem{Lenart2023}
A. Lenart et al. 2023, \emph{A bias-free cosmological analysis with quasars alleviating H0 tension}, ApJS, 264 46

\bibitem{ZhangMeszaros2001}
B. Zhang \& P. Mészáros 2001, \emph{Gamma-Ray Burst Afterglow with Continuous Energy Injection: Signature of a Highly Magnetized Millisecond Pulsar}, ApJ, 552 L35

\bibitem{DallOsso2011}
S. Dall'Osso et al. 2011, \emph{Gamma-ray bursts afterglows with energy injection from a spinning down neutron star}, A\&A, 526 A121

\bibitem{Rowlinson2013}
A. Rowlinson et al. 2013, \emph{Signatures of magnetar central engines in short GRB light curves}, MNRAS, 430 2 1061-1087

\bibitem{Bernardini2013}
M. G. Bernardini et al. 2013, \emph{How to switch on and off a Gamma-ray burst through a magnetar}, ApJ, 775 1 67

\bibitem{Stratta2018}
G. Stratta et al. 2018, \emph{On the Magnetar Origin of the GRBs Presenting X-Ray Afterglow Plateaus}, ApJ, 869 155

\bibitem{Rowlinson2014}
A. Rowlinson et al. 2014, \emph{Constraining properties of GRB magnetar central engines using the observed plateau luminosity and duration correlation}, MNRAS, 443 2 1779–1787

\bibitem{Rea2015}
N. Rea et al. 2015, \emph{Constraining the GRB-magnetar model by means of the Galactic pulsar population}, ApJ, 813 92

\bibitem{Srinivasaragavan2020}
G. P. Srinivasaragavan et al. 2020, \emph{On the Investigation of the Closure Relations for Gamma-Ray Bursts Observed by Swift in the Post-plateau Phase and the GRB Fundamental Plane}, ApJ, 903 18

\bibitem{Dainotti2021closure}
M. G. Dainotti et al. 2021, \emph{Closure relations during the plateau emission of Swift GRBs and the fundamental plane}, PASJ, 73 4 970–1000

\bibitem{Dainotti2023closure}
M. G. Dainotti et al. 2023, \emph{The Closure Relations in High-Energy Gamma-ray Bursts Detected by Fermi-LAT}, Galaxies, 11(1) 25

\bibitem{Fraija2022sinch}
N. Fraija et al. 2022, \emph{Synchrotron Self-Compton Afterglow Closure Relations and Fermi-LAT-detected Gamma-Ray Bursts}, ApJ, 934 188

\bibitem{Fraija2023closure}
N. Fraija et al. 2023, \emph{Off-axis Afterglow Closure Relations and Fermi-LAT Detected Gamma-Ray Bursts}, ApJ, 958 126

\bibitem{Dainotti2022optical}
M. G. Dainotti et al. 2022, \emph{The closure relations in optical afterglow of Gamma-Ray Bursts}, ApJ, 940 169

\bibitem{Levine2023radio}
D. Levine et al. 2023, \emph{Interpretation of radio afterglows in the framework of the standard fireball and energy injection models}, MNRAS, 519 3 4670–4683

\bibitem{Levine2022}
D. Levine et al. 2022, \emph{Examining Two-dimensional Luminosity–Time Correlations for Gamma-Ray Burst Radio Afterglows with VLA and ALMA}, ApJ 925 15

\bibitem{Dainotti2020platinum}
M. G. Dainotti et al. 2020, \emph{The X-Ray Fundamental Plane of the Platinum Sample, the Kilonovae, and the SNe Ib/c Associated with GRBs}, ApJ, 904 97

\bibitem{Pan2013}
Y. Pan et al. 2013, \emph{Minimum Accretion Rate for Millisecond Pulsar Formation in Binary System}, IAU Symposium, 290, 291-292

\bibitem{YuanBeloborodov}
Y. Yuan et al. 2022, \emph{Magnetar Bursts Due to Alfvén Wave Nonlinear Breakout}, ApJ, 933 174

\bibitem{GRB210905A}
A. Rossi et al. 2022, \emph{A blast from the infant Universe: the very high-z GRB 210905A}, A\&A, 665 A125

\bibitem{Amati2002}
L. Amati et al. 2002, \emph{Intrinsic spectra and energetics of BeppoSAX Gamma-Ray Bursts with known redshifts}, A\&A, 390 81

\bibitem{Yonetoku2004}
D. Yonetoku et al. 2004, \emph{Gamma-Ray Burst Formation Rate Inferred from the Spectral Peak Energy-Peak Luminosity Relation}, ApJ, 609 935

\bibitem{Tsvetkova2021}
A. Tsvetkova et al. 2021, \emph{The Konus–Wind Catalog of Gamma-Ray Bursts with Known Redshifts. II. Waiting-Mode Bursts Simultaneously Detected by Swift/BAT}, ApJ, 908 83

\bibitem{Ghirlanda2004}
G. Ghirlanda et al. 2004, \emph{The collimation-corrected GRB energies correlate with the peak energy of their $\nu F_\nu$ spectrum}, ApJ, 616 331-338

\bibitem{DainottiPetrosian2013}
M. G. Dainotti et al. 2013, \emph{Determination of the intrinsic Luminosity Time Correlation in the X-ray Afterglows of GRBs}, ApJ, 774 2 157

\bibitem{Petrosian2015}
V. Petrosian et al. 2015, \emph{Cosmological Evolution of Long Gamma-ray Bursts and Star Formation Rate}, ApJ, 806 1 44

\bibitem{AGILE}
M. Tavani et al. 2009, \emph{The AGILE Mission}, A\&A, 502 995-1013

\bibitem{Ursi2022b}
A. Ursi et al. 2022, \emph{The Second AGILE MCAL Gamma-Ray Burst Catalog: 13 yr of Observations}, ApJ, 925 152

\bibitem{Ursi2022c}
A. Ursi et al. 2022, \emph{AGILE Observations of GRB 220101A: A "New Year's Burst" with an Exceptionally Huge Energy Release}, ApJ, 933 2 214

\bibitem{Tavani2023}
M. Tavani et al. 2023, \emph{AGILE gamma-ray detection of the exceptional GRB 221009A}, ApJL, 956 1 L23 11

\bibitem{Ursi2022a}
A. Ursi et al. 2022, \emph{AGILE Observations of the LIGO-Virgo Gravitational-wave Events of the GWTC-1 Catalog}, ApJ, 924 2 80

\bibitem{Tavani2021}
M. Tavani et al. 2020, \emph{An X-Ray Burst from a Magnetar Enlightening the Mechanism of Fast Radio Bursts}, Nat. Astr., 5 401–407

\bibitem{Ursi2023ApJS}
A. Ursi et al. 2023, \emph{The First AGILE Solar Flare Catalog}, ApJS, 267 9

\bibitem{Ursi2020}
A. Ursi et al. 2020, \emph{AGILE and Konus-Wind Observations of GRB 190114C: The Remarkable Prompt and Early Afterglow Phases}, ApJ, 904 133

\bibitem{Ripa2023}
J. Ripa et al. 2023, \emph{The peak-flux of GRB 221009A measured with GRBAlpha}, A\&A 677, L2

\bibitem{GCN33424}
J. Ripa et al. 2023, \emph{GRB 230307A: VZLUSAT-2 detection}, GCN 33424

\bibitem{Bernardini2021}
M. G. Bernardini et al. 2021, \emph{The SVOM Mission}, 9(4) 113

\bibitem{MereghettiBalman}
S. Mereghetti et al. 2021, \emph{Time Domain Astronomy with the THESEUS Satellite}, submitted to Experimental Astronomy

\bibitem{Boller}
T. Boller 2023, \emph{Einstein Probe Mission}, Multifrequency Behaviour of High Energy Cosmic Sources - XIV, 12-17 June 2023

\bibitem{Hudec2022}
R. Hudec \& C. Feldman 2022, \emph{Lobster Eye X-ray Optics}, Handbook of X-ray and Gamma-ray Astrophysics

\bibitem{Bisnovatyi-Kogan}
G. S. Bisnovatyi-Kogan \& F. Giovannelli 2017, \emph{Time lag in transient cosmic accreting sources}, A\&A, 599 A55

\bibitem{Panessa2019}
F. Panessa et al. 2019, \emph{The origin of radio emission from radio-quiet active galactic nuclei}, Nat. Astr., 3 387–396

\bibitem{Sbarrato2022}
T. Sbarrato et al. 2022, \emph{Blazar nature of high–z radio–loud quasars}, A\&A, 663 A147

\bibitem{Shen2020}
X. Shen, X et al. 2020, \emph{The bolometric quasar luminosity function at $z=0-7$}, MNRAS, 495 3252

\bibitem{Belladitta2020}
S. Belladitta et al. 2020, \emph{The first blazar observed at $z>6$}, A\&A, 635 L7

\bibitem{Marshall}
H. L. Marshall et al. 2022, \emph{Observations of 4U 1626-67 with the Imaging X-ray Polarimetry Explorer}, ApJ, 940 70

\bibitem{Werner2022}
N. Werner et al. 2022, \emph{Quick Ultra-VIolet Kilonova surveyor (QUVIK)}, SPIE Astronomical Telescopes and Instrumentation

\bibitem{Abbasi2022}
R. Abbasi et al. 2022, \emph{Evidence for neutrino emission from the nearby active galaxy NGC 1068}, Science, 378 6619 538-543

\bibitem{Gasparri2022}
E. Gasparri et al. 2022, \emph{Search for Gamma-Ray counterparts of IceCube neutrino events in the AGILE public archive}, in prep.

\bibitem{Dolgov2023}
A. D. Dolgov, \emph{Tension between HST/JWST and $\Lambda$CDM Cosmology, PBH, and Antimatter in the Galaxy}, arXiv:2310.00671

\bibitem{DolgovSilk1993}
A. Dolgov \& J. Silk 1993, \emph{Baryon isocurvature fluctuations at small scales and baryonic dark matter}, Phys. Rev. D, 47 4244

\bibitem{Giovannelli2016}
F. Giovannelli \& L. Sabau-Graziati 2017, \emph{Frontier Research in Astrophysics: A Review}, Frontier Research in Astrophysics – II (FRAPWS2016), 269

\bibitem{Giovannelli2019}
F. Giovannelli \& L. Sabau-Graziati 2020, \emph{Multifrequency Behaviour of High Energy Cosmic Sources in the GW Era}, PoS MULTIF2019, 003

\bibitem{Erickson2007}
J. K. Erickson et al. 2007, \emph{Cosmic perturbations through the cyclic ages}, Phys. Rev. D, 75 123507

\bibitem{Arbuzova2022}
E. V. Arbuzova 2022, \emph{New Options for SUSY-Kind Dark Matter}, 	arXiv:2201.05127

\bibitem{Battistelli2022}
E. S. Battistelli et al. 2022, \emph{High angular resolution Sunyaev Zel'dovich observations: the case of MISTRAL}, arXiv:2204.04222

\bibitem{Orlando2023}
S. Orlando 2023, \emph{Modeling the Evolution from Massive Stars to Supernovae and Supernova Remnants}, arXiv:2311.05612

\bibitem{MiyamotoIshida}
A. Miyamoto et al. 2022, \emph{Understanding the physical state of hot plasma formed through stellar wind collision in WR140 using high-resolution X-ray spectroscopy}, MNRAS, 513 4 6074-6087

\bibitem{Merafina}
M. Merafina 2017, \emph{Dynamical evolution of globular clusters: Recent developments}, Int. J. Mod. Phys. D, 26 09 1730017

\bibitem{Kepler}
J. J. Lissauer et al. 2023, \emph{Exoplanet Science from Kepler}, Protostars and Planets VII, ASP Conference Series, Vol. 534

\end{thebibliography}
\end{document}